\documentclass[preprint]{aa}
\usepackage{graphics}

\begin{document}

%
%==================================================================%%
%%                                                                  %%
%%                                                                  %%
%%                                                                  %%
%%                      A S T R O N O M Y                           %%
%%                                                                  %%
%%                           AND                                    %%
%%                                                                  %%
%%                  A S T R O P H Y S I C S                         %%
%%                                                                  %%
%%                                                                  %%
%%        LaTeX Support                             Version 2.10    %%
%%                                                                  %%
%%==================================================================%%
%
%  Abbreviations
%
\def\etal {et al.}
\def\ie {i.\,e.}
\def\etseq {{\em et seq.}}
\def\vs {{it vs.}}
\def\perse {{it per se}}
\def\adhoc {{\em ad hoc}}
\def\eg {e.\,g.}
\def\etc {etc.}
\def\ccpers {\hbox{${\rm cm}^3{\rm s}^{-1}$}}
\def\DEGR {\hbox{$^{\circ }$}}
\def\vlsr {\hbox{${v_{\rm LSR}}$}}
\def\vel {\hbox{${v_{\rm LSR}}$}}
\def\vhel {\hbox{${v_{\rm HEL}}$}}
\def\delv {\hbox{$\Delta v_{1/2}$}}
\def\dvel {\hbox{$\Delta v_{1/2}$}}
\def\TL {$T_{\rm L}$}
\def\TC {$T_{\rm c}$}
\def\TEX {$T_{\rm ex}$}
\def\TMB {$T_{\rm MB}$}
\def\TKIN {$T_{\rm kin}$}
\def\TREC {$T_{\rm rec}$}
\def\TSYS {$T_{\rm sys}$}
\def\TVIB {$T_{\rm vib}$}
\def\TROT {$T_{\rm rot}$}
\def\TDUST {$T_{\rm d}$}
\def\TASTAR {$T_{\rm A}^{*}$}
\def\TVIBST {$T_{\rm vib}^*$} 
\def\TB {$T_{\rm B}$}
\def \la{\mathrel{\mathchoice   {\vcenter{\offinterlineskip\halign{\hfil
$\displaystyle##$\hfil\cr<\cr\sim\cr}}}
{\vcenter{\offinterlineskip\halign{\hfil$\textstyle##$\hfil\cr
<\cr\sim\cr}}}
{\vcenter{\offinterlineskip\halign{\hfil$\scriptstyle##$\hfil\cr
<\cr\sim\cr}}}
{\vcenter{\offinterlineskip\halign{\hfil$\scriptscriptstyle##$\hfil\cr
<\cr\sim\cr}}}}}
\def \ga{\mathrel{\mathchoice   {\vcenter{\offinterlineskip\halign{\hfil
$\displaystyle##$\hfil\cr>\cr\sim\cr}}}
{\vcenter{\offinterlineskip\halign{\hfil$\textstyle##$\hfil\cr
>\cr\sim\cr}}}
{\vcenter{\offinterlineskip\halign{\hfil$\scriptstyle##$\hfil\cr
>\cr\sim\cr}}}
{\vcenter{\offinterlineskip\halign{\hfil$\scriptscriptstyle##$\hfil\cr
>\cr\sim\cr}}}}}
\def\SDOZ {\hbox{$S_{12\mu \rm m}$}}
\def\STWE {\hbox{$S_{25\mu \rm m}$}}
\def\SSIX {\hbox{$S_{60\mu \rm m}$}}
\def\SHUN {\hbox{$S_{100\mu \rm m}$}}
\def\solmass {\hbox{M$_{\odot}$}}
\def\solum {\hbox{L$_{\odot}$}}
\def\irlum {\hbox{$L_{\rm IR}$}}
\def\ohlum {\hbox{$L_{\rm OH}$}}
\def\blum {\hbox{$L_{\rm B}$}}
\def\numd {\hbox{$n\,({\rm H}_2$)}}                   
\def\rhounit {$\hbox{M}_\odot\,\hbox{pc}^{-3}$}
\def\kms {\hbox{${\rm km\,s}^{-1}$}}
\def\kmsyr {\hbox{${\rm km\,s}^{-1}\,{\rm yr}^{-1}$}}
\def\kmsmpc {\hbox{${\rm km\,s}^{-1}\,{\rm Mpc}^{-1}$}} 
\def\Kkms {\hbox{${\rm K\,km\,s}^{-1}$}}
\def\percc {$\hbox{{\rm cm}}^{-3}$}    %cm-3
\def\cmsq  {$\hbox{{\rm cm}}^{-2}$}    %cm-2
\def\cmsix  {$\hbox{{\rm cm}}^{-6}$}  %cm-6
\def\arcsec {\hbox{$^{\prime\prime}$}}
\def\arcmin {\hbox{$^{\prime}$}}
\def\ffam {\hbox{$\,.\!\!^{\prime}$}}
\def\ffas {\hbox{$\,.\!\!^{\prime\prime}$}}
\def\ffM {\hbox{$\,.\!\!\!^{\rm M}$}}
\def\ffm {\hbox{$\,.\!\!\!^{\rm m}$}}
\def\ffs {\hbox{$\,.\!\!^{\rm s}$}}
\def\ffd {\hbox{$\,.\!\!^{\circ}$}}
\def\HI  {\hbox{HI}}
\def\HII {\hbox{HII}}

\def\kms{km\thinspace s$^{-1}$}
\def\cm2{cm$^{-2}$}

  \thesaurus{11     % A&A Section 11: Extragalactic Astronomy
              (09.13.2;  % ISM: molecules
               11.09.1;  % Galaxies: individual
               11.09.4;  % Galaxies: ISM
               11.19.3;  % Galaxies: starburst
               13.19.3)} % Radio lines: galaxies

\title{Giant Molecular Clouds in the Dwarf Galaxy NGC1569}

\author{C.L. Taylor\inst{1}
\and S. H\"uttemeister \inst{2}
\and U. Klein \inst{2}
\and A. Greve \inst{3}}

\institute{Ruhr-Universit\"at Bochum, Astronomisches Institut,
Universit\"atsstra\ss{}e 150, 44780 Bochum, Germany 
\and Universit\"at Bonn, Radioastronomisches Institut, Auf dem H\"ugel 71,
53121 Bonn, Germany
\and IRAM, 300 Rue de la Piscine, 38406 St. Martin d'H\`eres, France }

\offprints{C.L. Taylor}

\date{Received / Accepted }

\maketitle

\begin{abstract}
We present CO 1$\rightarrow$0 and 2$\rightarrow$1 observations of
the dwarf starburst galaxy NGC~1569 with the IRAM 
interferometer on Plateau de Bure.  We find the CO emission is 
not spatially associated with the two super star clusters in the
galaxy, but rather is found in the vicinity of an HII
region.  With the resolution of our data, we can resolve the
CO emission into five distinct giant molecular clouds, four
are detected at both transitions.  In the 1$\rightarrow$0 transition the
sizes and linewidths are similar to those of GMCs in the Milky Way
Galaxy and other nearby systems, with diameters ranging from
$\sim$ 40 to 50 pc and linewidths from 4 to 9 \kms. The (2-1)/(1-0) 
line ratios range from 0.64 $\pm$ 0.30 to 1.31
$\pm$ 0.60 in the different clouds. The lower line ratios are similar
to those seen in typical Galactic GMCs, while values higher than unity
are often seen in interacting or starburst galaxies. We use the virial 
theorem to derive the CO-H$_2$ conversion factor for three of the clouds, 
and we adopt an average value of 6.6 $\pm$ 1.5 times the Galactic 
conversion factor for NGC~1569 in general.  We discuss the role of the 
molecular gas in NGC~1569, and its relationship to the hot component of 
the ISM.  Finally, we compare our observations with blue compact 
dwarf galaxies which have been mapped in CO.

\end{abstract}

\section{Introduction}

NGC1569 (Arp 210, VII~Zw~16, UGC3056) is a nearby dwarf galaxy hosting
several interesting phenomena related to its starburst, and with an 
observational history going back to 1789 (see Israel \cite{I88} for a history
of the early observations).  As is common among dwarf galaxies, it
has a low metallicity (12 + log(O/H) = 8.19 $\pm$ 0.02; Kobulnicky
\& Skillman \cite{KS97}). At a distance of only 2.2 $\pm$ 0.6 Mpc
(Israel \cite{I88}), it is the closest known example of a dwarf starburst
galaxy, and so observations of it are essential in interpreting 
observations of similar objects at greater distances (e.g. the small
blue galaxies found in the Hubble Deep Fields).

What has perhaps drawn the greatest attention to NGC1569 is the 
presence of two super star clusters (SSCs), labeled A and B 
(Ables \cite{A71} , Arp \& Sandage \cite{AS}).  These clusters have been the
subject of recent HST (O'Connell, Gallagher \& Hunter \cite{OGH}; De Marchi
et al. \cite{DM}) and ground based studies (Prada et al. \cite{PGM}; Ho 
\& Filippenko \cite{HF}; Gonzalez Delgado et al. \cite{GD}), and are 
believed to be similar to young globular clusters.  Age estimates for the 
SSCs range from 3 to 10 Myr depending upon assumptions about the star 
formation history.

Greggio et al. (\cite{GTe98}) have determined from  HST WFPC2 images 
that NGC~1569 has experienced a global burst of star formation 100
Myr in duration that has ended as recently as 5 Myr ago.  Vallenari \& 
Bomans (\cite{VB}) found evidence for a large burst of star formation 
roughly 1$~\times~10^8$ yr ago in WFPC images, as well 
as several much older episodes.
It is clear that star formation has had a dramatic effect upon the
ISM in NGC1569 through stellar winds and supernovae.  Israel \& van Driel
(\cite{IvD}) have found a hole in the HI distribution centered on SSC A, 
possibly blown out by the stars in the cluster.  H$\alpha$ emission
extends from the disk out to the halo in filamentary structures (Hodge
\cite{H74}; Waller \cite{W91}; Hunter, Hawley \& Gallagher \cite{HHG}; 
Tomita, Ohta \& Saito \cite{TOS}; Devost, Roy \& Drissen 
\cite{DRD}).  The dynamical age of the
extended, diffuse H$\alpha$ emission is consistent with some age
determinations of the SSCs (Heckman et al. \cite{HDL}).  X-ray studies
(Heckman et al. \cite{HDL}; Della Ceca et al. \cite{DCG}) have found
extended emission spatially associated with the H$\alpha$; together
these paint a picture of a hot gaseous phase blowing out of the galaxy,
powered by the SSCs.

The cool phase of the ISM in NGC 1569 has been observed by
Israel \& van Driel (\cite{IvD}), Stil \& Israel (\cite{SI}) and Wilcots 
et al. (in preparation) in HI with interferometers. Hunter et al. 
(\cite{HTCH}) have combined FIR data from the Kuiper Airborne Observatory 
and IRAS to study the dust.  They find an unusually high dust temperature,
for a dwarf irregular galaxy, 34\,K, and attribute this to the influence
of the recent star formation burst.   CO emission has been observed in 
NGC1569 by Young, 
Gallagher \& Hunter (\cite{YGH}), Greve et al. (\cite{GBJM}) and
Taylor, Kobulnicky \& Skillman (\cite{TKS}).  Mapping with the
IRAM 30-m telescope, Greve et al. found CO emission near the SSCs,
but not directly at their locations.  Instead the CO is spatially
associated with a prominent HII region.  The spatial resolution of
the 30-m telescope is 22\arcsec\ at 115 GHz (equivalent to 235 pc) 
and 11\arcsec\ at 230 GHz (equivalent to 120 pc), so it was not 
possible to resolve 
individual molecular clouds.  Because of this, an extremely high value
of the CO-to-H$_2$ conversion rate, $\sim$ 20 times the Galactic value
was derived.  Such a high value is unlikely, and Greve et al. 
attributed it to non-standard conditions in the molecular gas due to
the proximity of the super star clusters.

NGC~1569 presents an excellent opportunity to study the effects of 
low metallicity and a  starburst upon the molecular ISM. 
The only similar environment which can be studied in such  detail
is the molecular gas complex found near 30 Doradus in the LMC.
We have imaged the CO emission using the IRAM interferometer
on Plateau de Bure, which provides spatial resolution sufficient to 
distinguish individual giant molecular clouds (GMCs).  We examine
the physical properties of the GMCs, derive the CO-to-H$_2$ conversion
rate, and discuss the role of the molecular ISM in NGC1569 with respect
to its star formation history.

\section{Observations and Data Reduction}

\begin{figure}
\resizebox{\hsize}{!}{\includegraphics{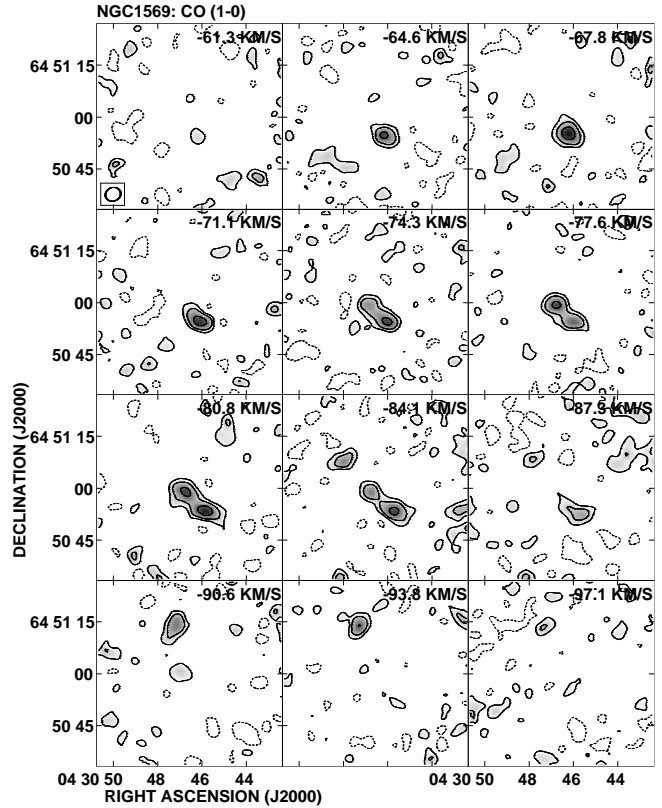}}
\caption{Channel maps showing $^{12}$CO 1$\rightarrow$0 emission in
NGC~1569 near SSC A and B.  The contours are in units of -2, 2, 4, and 
8$\sigma$, where $\sigma$ is the rms noise in a channel map, equal to 7.5 
mJy beam$^{-1}$.  Every second channel is shown.  Velocities are v$_{LSR}$.}
\label{fig1}
\end{figure}

\begin{figure}
\resizebox{\hsize}{!}{\includegraphics{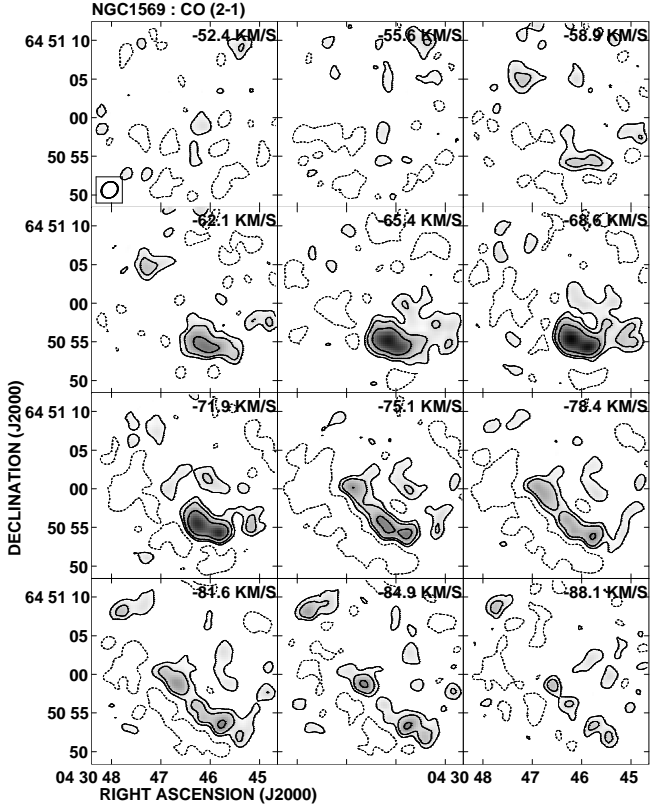}}
\caption{Channel maps showing $^{12}$CO 2$\rightarrow$1 emission in
NGC~1569 near SSC A and B.  The contours are in units of -2, 2, 4, and 
8$\sigma$, where $\sigma$ is the rms noise in a channel map, equal to 6.6 
mJy beam$^{-1}$.  Velocities are v$_{LSR}$.}
\label{fig2}
\end{figure}

The observations were carried out with the interferometer on 
Plateau de Bure in several sessions between 21 March and 14 April
1998.  The $^{12}$CO 1$\rightarrow$0 and 2$\rightarrow$1 transitions 
were observed simultaneously in a mosaic consisting of four positions 
separated by 11\arcsec , centered upon the peak in the CO emission 
detected by Greve et al. (\cite{GBJM}).  The D and C2 configurations 
were used, yielding spatial resolutions of 4\ffas 5 $\times$ 
3\ffas 9  at 115 GHz and 2\ffas 3 $\times$ 2\ffas 0  
at 230 GHz.  At the adopted distance of 2.2 Mpc, these spatial 
resolutions correspond to 45 pc $\times$ 40 pc and 25 pc $\times$ 20 pc,
respectively The velocity resolutions are 1.6 \kms\ at 115 GHz and 3.3 \kms\
at 230 GHz.

The data were reduced at IRAM with the
CLIC and MAPPING packages of GILDAS, using the standard procedures.
During the observations the atmospheric water vapor content varied,
and was often high enough to influence the 230 GHz observations.  
The phase calibration was accomplished accurately with the help of
the stable phases at 115 GHz, but the amplitude calibration at
230 GHz is only accurate to 40\%.  The accuracy of the 115 GHz amplitude
calibration is 20\%.  After mapping was completed the data cubes 
were exported as FITS files to the AIPS package for further analysis.

The 115 GHz data have an rms noise of 7.5 mJy/beam in a single
channel map, while for the 230 GHz data the rms noise is 6.6 mJy/beam.
The channel maps are presented in Figures 1 and 2.

%\begin{figure*}
%\resizebox{12cm}{!}{\includegraphics{fig1.ps}}
%\caption{Channel maps showing $^{12}$CO 1$\rightarrow$0 emission in
%NGC~1569.  The contours are in units of -2, 2, 4, and 8$\sigma$, where
%$\sigma$ is the rms noise in a channel map, equal to 7.5 mJy beam$^{-1}$.}
%\label{fig1}
%\end{figure*}

Each data cube was blanked at the 3$\sigma$ level and the resulting
blanked data cubes were searched for CO emission.  To distinguish 
genuine emission from noise spikes, additional blanking
was done in which only emission present in at least three consecutive
channels was retained.  The integrated CO emission over the 
entire map is 3.58 Jy \kms\/ in the 1$\rightarrow$0 line, and 5.94 Jy 
\kms\/ in the 2$\rightarrow$1 line.  In comparison, Greve et al. (\cite{GBJM}) 
integrated the emission over the inner
22\arcsec\ of their map, obtaining 12.5 Jy \kms\/ (1.98 K \kms) and
22.1 Jy \kms\/ (2.10 K \kms) for the 1$\rightarrow$0 line and 
2$\rightarrow$1 line, respectively.  Our field of view is larger than
this, so if we restrict ourselves to the same inner 22$^{\arcsec}$, we
obtain 2.76 Jy \kms\/ and 5.79 Jy \kms.  At 115 GHz, we detect $\sim$
22\% of the flux from Greve et al., and at 230 GHz we detect $\sim$ 26\%.
This discrepancy is most likely due to the incomplete coverage of the {\it uv}
plane at the shortest baselines.   To check that insufficient sensitivity
was not the reason, we assumed that the entire 22\arcsec\ region in our 
1$\rightarrow$0 map, excluding only the areas covered by the GMCs, was 
filled with emission 3 channels wide at the 2.5$\sigma$ level, i.e. just 
below our sensitivity defined above.  If this were true, then the total 
emission should be 3.82 Jy \kms, and we would have recovered only 72\% of 
this.  But even this most extreme scenario still falls short of the 12.5 
Jy \kms\/ seen by Greve et al.  Thus we conclude that the discrepancy is 
due to the missing short spacings.  This lack of short spacings
means that our observations do not detect diffuse gas which is distributed
over large scale lengths, but only the dense gas which has accumulated
into giant molecular clouds.  Our results will therefore only be applicable
to the structures we see, and not to the global distribution of CO in 
NGC1569.

\section{Results}

\subsection{Cloud Diameters, Line Widths and Line Ratios}

\begin{table*}[t]
\begin{center}
{\bf Table~1}: Observed Properties of GMCs in NGC1569
\end{center}
\begin{center}
\begin{tabular*}{14cm}{lccccccc}
\hline
% GMC no. & $\alpha$(2000) & $\delta$(2000) & Diameter(pc) & Diameter (pc) & FWHM (km/s) & V(center) & I(CO) Jy km/s
GMC & $\alpha$(2000) & $\delta$(2000) &D$_{90}$ & D$_{fwhm}$ & v$_{fwhm}$ & v$_{center}$ & S(CO) \\
 & & & (pc) & (pc) & (\kms) & (\kms) & (Jy \kms) \\
\hline
{\it 1$\rightarrow$0} & & & & & \\
1+2 & 04 30 46.1 & +64 50 55 & 110 $\times$ 57 & 70 $\times$ 44 & 24.1 & --75.7 & 2.12 \\
3 & 04 30 46.6 & +64 50 58 & 72 $\times$ 57 & 61 $\times$ 35 & 8.8 & --79.3 & 0.86 \\
4 & 04 30 47.9 & +64 51 08 & 75 $\times$ 45 & 55 $\times$ 33 & 3.9 & --85.0 & 0.19 \\
5 & 04 30 47.2 & +64 51 14 & 81 $\times$ 56 & 55 $\times$ 37 & 5.7 & --92.1 & 0.41 \\
\hline
{\it 2$\rightarrow$1} & & & & & \\
1 & 04 30 45.8 & +64 50 54 & 41 $\times$ 34 & 31 $\times$ 24 & 11.3 & --68.7 & 1.98 \\
2 & 04 30 46.1 & +64 50 55 & 47 $\times$ 38 & 37 $\times$ 28 & 13.1 & --69.1 & 2.28 \\
3 & 04 30 46.7 & +64 50 59 & 49 $\times$ 26 & 36 $\times$ 21 & 11.4 & --79.9 & 0.83 \\
4 & 04 30 47.8 & +64 51 08 & 28 $\times$ 21 & 24 $\times$ 16 & 4.9 & --84.9 & 0.23 \\
\hline
\end{tabular*}
\vbox{\hsize 13cm
{\it Note to Table 1}: The estimated errors on the various properties in the
CO 1$\rightarrow$0 data are: diameter $\pm$ 24 $\times$ 21 pc, v$_{fwhm}$
$\pm$ 1.3 \kms, v$_{center}$ $\pm$ 1.3 \kms, I(CO) $\pm$ 20\%.  For the 
CO 2$\rightarrow$1 data the estimated errors are: diameter $\pm$ 13 $\times$ 11 
pc,  v$_{fwhm}$ $\pm$ 3.2 \kms, v$_{center}$ $\pm$ 3.2 \kms, I(CO) $\pm$ 40\%.
}
\end{center}
\end{table*}

The zeroth moment maps, showing the velocity-integrated emission, are 
given in Figures 3 and 4.  Figure 5 shows the line profiles
for the clouds in both CO 1$\rightarrow$0 and 
2$\rightarrow$1 emission.  Although the CO emission from 
some giant molecular clouds overlaps spatially, the velocity information
shows that several distinct clouds may be distinguished, especially in
the 2$\rightarrow$1 data, which have a higher spatial resolution.  
Four clouds
are identified in each data cube, although clouds 1 and 2 from the
2$\rightarrow$1 are merged together by the low spatial resolution in
the 1$\rightarrow$0 transition.  Cloud 5 from the 1$\rightarrow$0 data
falls outside the area covered in the 2$\rightarrow$1 map and thus is not
seen. The observed properties of the GMCs are given in Table~1, including
the diameter measured at the contour encircling 90\% of the flux
(used for determining the virial masses), the diameter measured at the 
half maximum contour (used for comparing with the size-linewidth relation), 
the FWHM velocity width, the central velocity, 
and the integrated intensity.

\begin{figure}
\resizebox{\hsize}{!}{\includegraphics{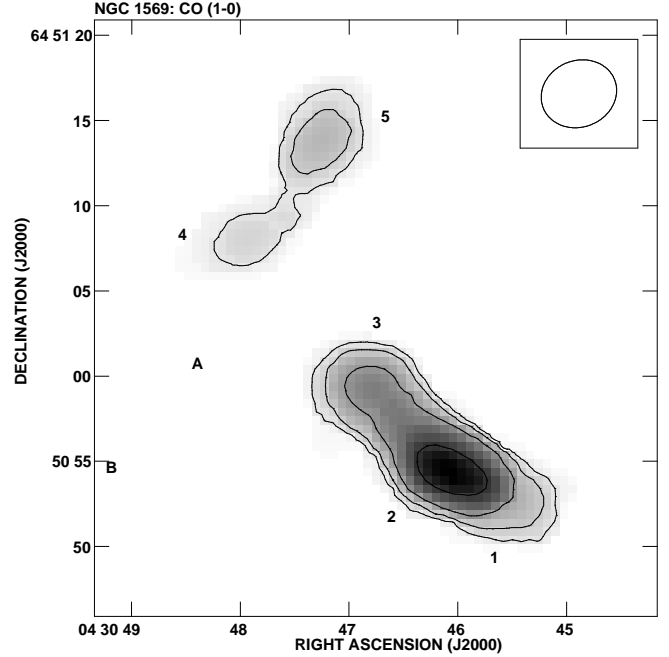}}
\caption{Integrated CO intensity map of the $^{12}$CO 1$\rightarrow$0
in NGC~1569.  The contours represent 10, 20, 40 and 80\% of the peak
integrated flux intensity, equal to 1.6 Jy \kms.  The positions of 
the two SSCs are labeled A and B.  The ellipse indicates
the size and shape of the synthesized beam.}
\label{fig3}
\end{figure}

\begin{figure}
\resizebox{\hsize}{!}{\includegraphics{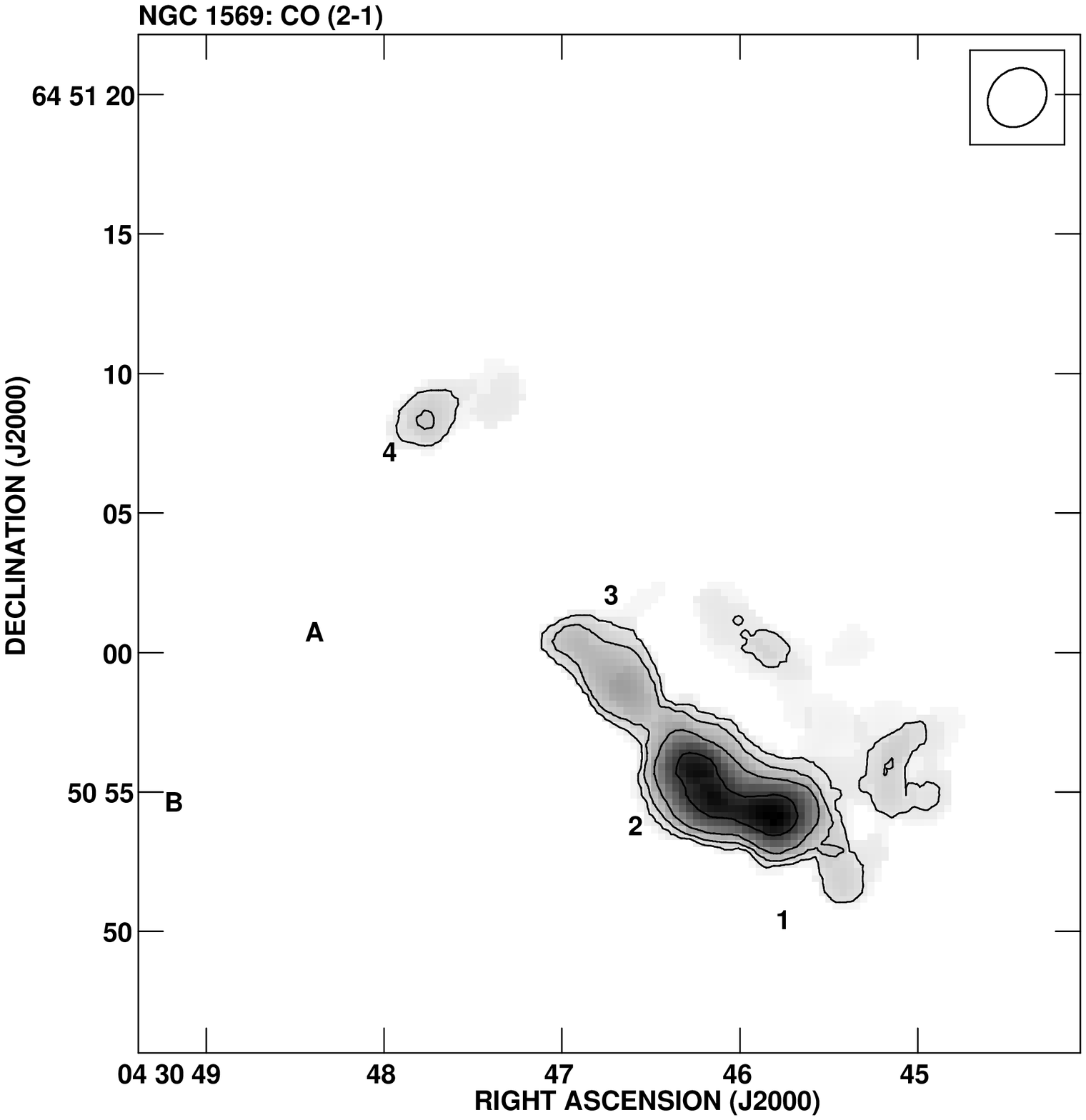}}
\caption{Integrated CO intensity map of the $^{12}$CO 2$\rightarrow$1
in NGC~1569.  The contours represent 10, 20, 40 and 80\% of the peak
integrated flux intensity, equal to 1.8 Jy \kms. The positions of 
the two SSCs are labeled A and B.   The ellipse indicates
the size and shape of the synthesized beam.}
\label{fig4}
\end{figure}

\begin{figure}
\resizebox{\hsize}{!}{\includegraphics{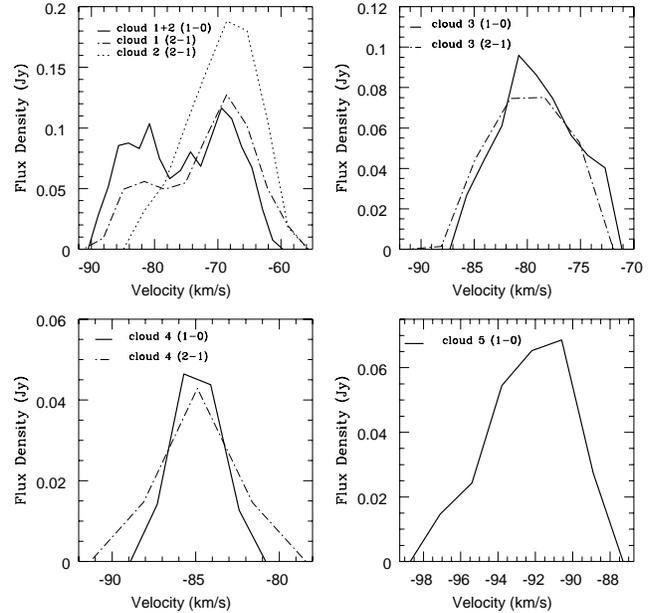}}
\caption{Line profiles for GMCs 1, 2, 3, 4 and 5.  Clouds 1 and 2 are
blended together in the $^{12}$CO 1$\rightarrow$0 data because of the
lower spatial resolution.  Cloud 5 is only shown in the $^{12}$CO 
1$\rightarrow$0 transition because it lies outside the area imaged
at 230 GHz.}
\label{fig5}
\end{figure}

One question to consider is how these GMCs in NGC1569 compare with
those from the Milky Way Galaxy, and from other nearby galaxies.  Both in the
Milky Way and other galaxies, GMCs are observed to follow a size-linewidth
relation of the form $v \propto D^{\beta}$ where $v$ is the linewidth, $D$
the diameter, and $\beta \sim$ 0.5 (e.g. Larson \cite{L81}, Solomon et al. 
\cite{Se87}, Wilson \& Scoville \cite{WS90}).
Figure~6 plots our clouds in the size-linewidth plane along with clouds 
observed in M31 (Vogel et al. \cite{VBB}; Wilson \& Rudolph \cite{WR}),
M33 (Wilson \& Scoville \cite{WS90}), SMC (Rubio et al. \cite{RLB}), 
IC10 (Wilson \cite{W95}) and NGC~6822 (Wilson \cite{W94}).  The line shows 
a fit to the points from M33 by Wilson \& Scoville (\cite{WS90}) of the form: 
$v = 1.2 D^{0.5}$.  Our clouds clearly fall within the range spanned by 
the clouds from other galaxies.  

\begin{figure}
\resizebox{\hsize}{!}{\includegraphics{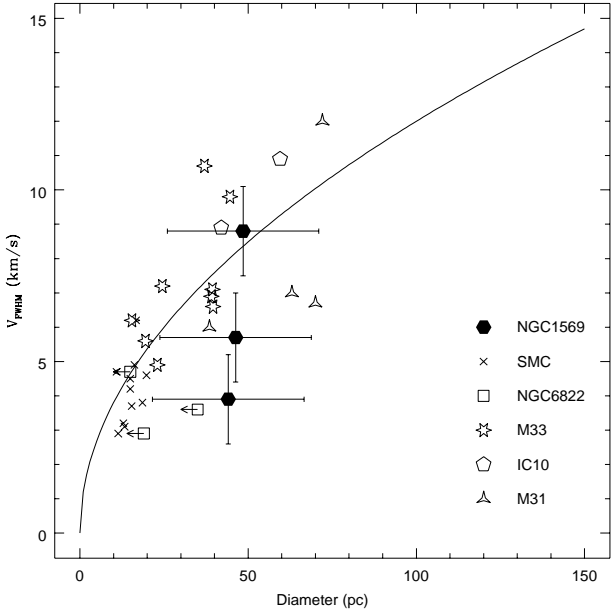}}
\caption{The diameter-line width relationship for GMCs in galaxies of the Local
Group.  The clouds of different galaxies are distinguished by different
symbols, and the line shows a best fit relationship to the M33 GMCs derived
by Wilson \& Scoville (\cite{WS90}).}
\label{fig6}
\end{figure}

The CO (2$\rightarrow$1)/(1$\rightarrow$0) line ratio, $r_{21}$, is given
in Table~2.  Greve et al.\ (\cite{GBJM}) determined the 
line ratios
for the inner 22\arcsec\ of their single dish map, obtaining 1.1 $\pm$
0.2.  This value lies between the extremes that we have determined for the 
individual GMCs, although our error bars are large enough to include
their value.  If we integrate only the emission corresponding to the 
same region as Greve et al., we obtain a line ratio of 1.38, larger 
than their value, but consistent, given our errors.  However, GMCs 1+2
contain most of the emission at both frequencies, so it is likely that
the value of Greve et al. is dominated by the combined contribution
of these two objects, and that the lower line ratios we obtain for 
GMCs 3 and 4 are correct.  

\begin{table}
\begin{center}
{\bf Table~2}: CO (2$\rightarrow$1)/(1$\rightarrow$0) Line Ratios
\end{center}
\begin{center}
\begin{tabular}{lc}
\hline
% GMC no. & line ratio
GMC & 2$\rightarrow$1/1$\rightarrow$0 \\
\hline
1+2 & 1.31 $\pm$ 0.60 \\
3 & 0.64 $\pm$ 0.30 \\
4 & 0.79 $\pm$ 0.36 \\
\hline
\end{tabular}
\end{center}
\end{table}

\subsection{The CO-H$_2$ Conversion Factor}

Under the assumption that the clouds we observe are in virial 
equilibrium, we may use the virial theorem to calculate the mass within
the clouds.  Since the mass of GMCs is dominated by molecular
hydrogen, comparing the mass thus obtained from the observed
CO flux density will give the conversion factor between CO and H$_2$.
For this calculation to apply, the clouds {\it must} be resolved both 
spatially and in velocity, otherwise upper limits on the molecular mass
of the GMCs are obtained, resulting in lower limits on the conversion
factor.  Figure~5 shows that the velocity resolution is sufficient, although
we will not be able to use the CO 1$\rightarrow$0 data for clouds 1 and 2
because they are blended together at the lower spatial resolution.  At a
distance of 2.2 Mpc, the spatial resolution at 115 GHz of 4\ffas 5 
$\times$ 3\ffas 9 corresponds to 48.0 $\times$ 41.6 pc, while 
the resolution at 230 GHz of 2\ffas 3 $\times$ 2\ffas 0  
is equal to 24.5 $\times$ 21.3 pc.

An important question is whether or not the assumption of virialization
is justified.  From Figure~6 we conclude that the clouds are similar to 
GMCs known in the Galaxy and other nearby systems, so if those clouds 
are virialized, we may reasonably assume that ours are as well.  That GMCs 
in the Milky Way are virialized has often been
the subject of vigorous debate.  One line of reasoning that argues for
virialized clouds has been the tight correlation between virial masses, 
M$_{VT}$, and CO luminosities, L$_{CO}$, where L$_{CO}$ is taken as an
indicator of the mass of CO, and hence H$_2$, present in a given cloud
(Solomon et al. \cite{Se87}).
However, Maloney (\cite{M90}) has argued that this M$_{VT}$--L$_{CO}$ 
correlation is simply a result of the observed size-linewidth relation
and would exist whether or not the GMCs were in virial equilibrium.
In the end, the fact that conversion factors derived with this method
generally agree with those derived from independent methods, at least
at the high mass end of the GMC distribution ($\geq 10^5$ M$_{\sun}$), 
suggests that the assumption of virialization is reasonable (Combes 
\cite{C91}).  Both sensitivity and spatial resolution limit our observations
to these largest, most massive GMCs, so we will use the viral masses
to derive the H$_2$-CO conversion factor.

The virial mass contained in a cloud is given by 

\begin{equation}
\label{eq1}
\hfill M_{VT} = 190\ \frac{v^2_{fwhm}}{{\rm km\,s^{-1}}}\
\frac{D/2}{\rm pc}\ \solmass \hfill
\end{equation}

\noindent
where $v^2_{fwhm}$ is the velocity width and $D$ the diameter of the
cloud (MacLaren et al. 
\cite{MRW}).  For the diameter, we use the average of the major and
minor axes, measured at the contour containing 90\% of the flux (D$_{90}$),
and for the velocity width, we use the linewidths in the 1$\rightarrow$0
line.  The factor of 190 is appropriate for a spherical 
distribution with density proportional to $1/r$.  The molecular mass in 
a given cloud of integrated CO 1$\rightarrow$0 flux density $S_{CO}$ 
[Jy \kms ] is: 

\begin{equation}
\label{eq2}
\hfill M_{mol} = 1.23~\times~10^4\ \frac{d^2}{\rm Mpc}\
\frac{S_{\rm CO}}{\rm Jy\,km\,s^{-1}}\ \solmass \hfill
\end{equation}

\noindent
where $d$ is the distance to the cloud (Wilson \& Scoville 
\cite{WS90}).  This formula uses a Galactic conversion factor 
of $\alpha_{Gal}$ = 2.3~$\times$~10$^{20}$ \cm2 (K \kms)$^{-1}$ 
(Strong et al. \cite{Se88}) and includes the helium correction. 
The CO-H$_2$ conversion factor is then obtained from 

\begin{equation}
\label{eq3}
\hfill \alpha = \alpha_{Gal} \frac{M_{VT}}{M_{mol}} \hfill
\end{equation}

Table~3 lists the values for $\alpha / \alpha_{gal}$ and
$M_{VT}$ for the three clouds with adequate spatial resolution 
to measure the diameters.  Also included are the estimated
masses of H$_2$.  For clouds 1+2 this is obtained using the
average conversion factor described below, for clouds 3, 4 and 5 
this is simply the virial mass reduced by a factor of 1.36 for
the helium contribution.

\begin{table}
\begin{center}
{\bf Table~3}: CO-H$_2$ Conversion Factors and Cloud Masses
\end{center}
\begin{center}
\begin{tabular}{lccc}
\hline
% GMC no. & conversion factor & M(VT) & M(H2)
GMC & $\alpha / \alpha_{gal}$ & $M_{VT}$ & M$_{H_2}$ \\
 & & $10^5$ M$_{\sun}$ & $10^5$ M$_{\sun}$ \\
\hline
1+2 & ... & ... & 10.9 $\pm$ 3.8 \\
3 & 7.2 $\pm$ 3.6 & 4.8 $\pm$ 1.8 & 3.5 $\pm$ 1.3 \\
4 & 5.9 $\pm$ 3.5 & 0.87 $\pm$ 0.44 & 0.64 $\pm$ 0.32 \\
5 & 6.7 $\pm$ 3.5 & 2.1 $\pm$ 0.85 & 1.5 $\pm$ 0.63 \\
\hline
\end{tabular}
\end{center}
\end{table}

The values for the individual clouds are consistent with each other, given 
the rather large error bars.  The scatter between the values, $\sigma$, is 
0.6, smaller than the individual errors, and smaller than the 20\% 
accuracy of the calibration of the CO flux.  Combining these two 
contributions, we adopt 1.5 as the error 
for the {\it average} conversion factor in NGC~1569 instead of using
the statistical scatter of the three values, thus $\alpha / \alpha_{gal}$ 
= 6.6 $\pm$ 1.5.   We can compare this with conversion factors calculated 
with the same method by Wilson (\cite{W95}) for other dwarf galaxies of 
nearly the same metallicity as NGC~1569.  NGC~6822, with 12 + log(O/H) = 
8.20, has $\alpha / \alpha_{gal}$ $<$ 2.2 $\pm$ 0.8, while IC10, with 12 
+ log(O/H) = 8.16 has $\alpha / \alpha_{gal}$ = 2.7 $\pm$ 0.5.  Thus 
NGC~1569 has a higher conversion rate than dwarf galaxies of similar 
metallicity by a factor or 2 to 3.  

Israel (\cite{I97}) has used far infrared data from IRAS to determine
the conversion factor in several magellanic irregular galaxies, 
including NGC1569.  He obtains $\alpha / \alpha_{gal}$ = 70 $\pm$ 35
for NGC1569, nearly ten times our value.  At least two possibilities
exist to explain this discrepancy.  Our value is only valid for those
GMCs in which we detect CO emission, while there may exist regions 
containing molecular material, but no CO.  Thus Israel's value may
be an average over the whole galaxy, including areas where there
is no CO emission.  Alternately, one or more of the assumptions used
by Israel in his determination of the conversion factor may not be
valid.  For example, his method assumes a constant dust-to-gas ratio
everywhere in the galaxy, which may not be the case.  Too little is
currently known about the distribution of cool dust in dwarf galaxies.

Using our derived conversion factor, the total mass of H$_2$ in the five 
detected GMCs is (16.5~$\pm$~4.1)~$\times$~10$^5$ M$_{\sun}$, compared
to 1.1~$\times~10^8$ M$_{\sun}$ in HI (Reakes \cite{R80}). The mass
of a typical single HI clump that can be identified in the map of 
Israel \& van Driel (1990) is of order $10^6$\,\solmass , i.e.\ similar to
the mass of the GMCs we find.
For comparison, we can use the absolute magnitude (M$_B$ = --16.9; 
Tully \cite{T88}) and an assumed stellar mass-to-light ratio of 1 to 
estimate the total stellar mass to be $\sim 10^9$ M$_{\sun}$.  
This mass-to-light ratio may be an
underestimate, because NGC~1569 has had a recent burst of star formation,
but we see that the stellar mass is much larger than the contribution made
by the molecular or atomic ISM.

Greve et al. (1996) have not found any CO emission in other regions in 
NGC~1569, but even if there was some they missed, it is unlikely to
account for much more than what is already seen.  We conclude that
the molecular gas contributes a small fraction by mass to the ISM
in NGC~1569.  {\em However, it should be noted that the conversion factor
we derive should only be applied in a strict sense to the GMCs.  The 
diffuse emission not detected by our interferometric observations may
have a different conversion factor, since it will have a different (lower)
density, and possibly a different temperature.}  

We can obtain an order of magnitude estimate of the density of this
diffuse gas.  Greve et al. argued that the total molecular mass they
detected is of the order 2~$\times$~10$^6$ M$_{\sun}$ for an assumed
conversion ratio of $\sim$ 4 times the Galactic value.  This should 
give a lower limit to the density, because for diffuse gas self shielding
of CO is less efficient and CO will be more easily dissociated.  Thus a
given mass of H$_2$ will have a higher conversion factor.  The interferometer
detects 22\% of the 1$\rightarrow$0 found by Greve et al., leaving a
mass of 1.6~$\times$~10$^6$ M$_{\sun}$ undetected.  For a spherical
volume of 120 pc radius (Greve et al.), this corresponds to an average 
density of $\sim$ 10 molecules cm$^{-3}$.

\section{Discussion}

\subsection{The Physical Conditions of the Molecular Gas}

Line ratios are often used as indicators of the physical conditions in
the molecular ISM.  We will compare the $^{12}$CO 2$\rightarrow$1 / 
1$\rightarrow$0 line ratios we present in Table~2 with those found for 
molecular gas in various environments. Sakamoto et al. (\cite{Se94}) have 
derived this line ratio for the GMCs Orion A and B, obtaining 0.77 and 
0.66 respectively, similar to our values for GMCs 3 and 4.  Thornley \& 
Wilson (\cite{TW}) have obtained a line ratio of 0.67 $\pm$ 0.19, averaged 
over several GMCs in M33.  Line ratios significantly higher than unity are
often considered indicative of warm, optically thin gas and have been
observed in interacting galaxies, or in galaxies with nuclear starbursts
(Braine \& Combes \cite{BC}; Aalto et al. \cite{Ae95}).  For the 
starburst system M82, Wild et al. (\cite{Wi92}) find a line ratio of 
1.3 $\pm$ 0.3.  In NGC1569
the GMCs with the highest line ratios (GMCs 1+2) are not significantly
nearer to either the SSCs or the closest HII regions in projected separation
than the other clouds, so it is unlikely that warming due to star
formation is responsible.  More likely the line ratios reflect a contrast
in the densities of the GMCs.

Because we have data only in the $^{12}$CO 1$\rightarrow$0 and 
2$\rightarrow$1 transitions, we cannot provide tight constraints
on the physical conditions in the CO emitting gas.  Generally a
line known to be optically thin is necessary for this, such as
$^{13}$CO 1$\rightarrow$0.  Given the low emission in
the lines of the much more common $^{12}$CO, obtaining $^{13}$CO
detections is difficult in dwarf galaxies.

However, we can run a series of large velocity gradient (LVG) models
and see what ranges of parameter space are consistent with our
observed 2$\rightarrow$0/1$\rightarrow$0 line ratios.  As input
parameters into the models we have a $^{12}$CO/H$_2$ abundance ratio
of 2~$\times~10^{-5}$ and velocity gradients of 0.2, 0.4, 1.0 and
2.0 km s$^{-1}$\,pc$^{-1}$. The abundance ratio is simply the standard 
assumption scaled by 0.2, as the metallicity of NGC1569 is
approximately 20\% solar.  

Because we do not have a third line, such as $^{13}$CO 1$\rightarrow$0,
the kinetic temperature T$_{\rm kin}$ and the density, n$_{\rm H_2}$ are 
degenerate and a large number of solutions may give the same line ratios
by trading off temperature versus density.  Figure~7 shows the result
of one LVG model, where the velocity gradient is 1.0 km s$^{-1}$\,pc$^{-1}$.
For this set of parameters, we see that the gas with the high line
ratio will have a narrow set of acceptable values for kinetic temperature 
($\sim$ 150 K), and densities (log n$_{\rm H_2}$ $\sim$ 3.5).  For the low 
line ratio gas, a wide range of acceptable values exists.  

\begin{figure}
\resizebox{\hsize}{!}{\includegraphics{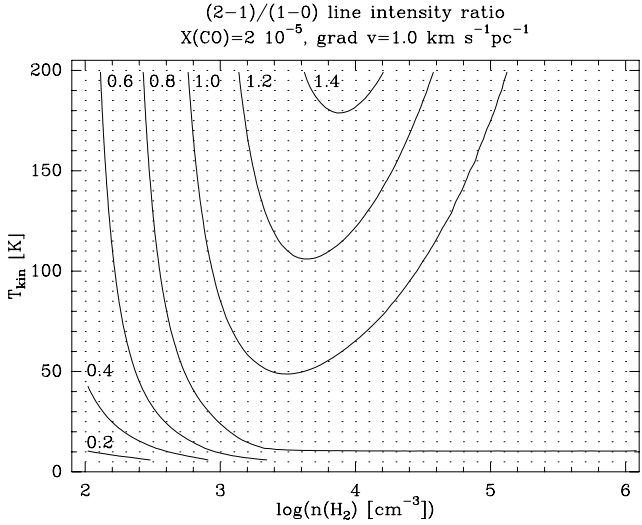}}
\caption{Results of an LVG model with $^{12}$CO/H$_2$ = 2~$\times~10^{-5}$
and a velocity gradient of 1 km s$^{-1}$ pc$^{-1}$.}
\label{fig7}
\end{figure}

The physical conditions in the molecular gas determine the value of the
CO--H$_2$ conversion factor.  Because the 1$\rightarrow$0 line is optically
thick under the conditions found in GMCs, the empirical relationship
that permits it to be used as a tracer of H$_2$ mass depends on a
very clumpy molecular medium with a low filling factor, where the clumps
do not shadow one another.  If this is not the case in NGC~1569 and there is
substantial shadowing, then this would reduce the derived $M_{mol}$, which
would, in turn, increase the conversion factor. 

Alternately, if the
GMCs are in a stronger UV radiation field than those in either IC10 or
NGC~6822, the CO emission could be reduced due to increased photo-dissociation
of the CO molecule relative to those two galaxies.  This would also
increase the conversion factor.  This explanation is consistent with
the different natures of these three galaxies.  NGC~1569 is a BCD
which has had a major burst of star formation in the recent past,
which resulted in the two SSCs.  The presence of HII regions also indicates
that some star formation is currently proceeding.  Both IC10 and NGC~6822
are far more quiescent than NGC~1569, although a high concentration of
Wolf-Rayet stars in IC10 does indicate a recent star formation
episode in that system (Massey \& Armandroff \cite{MA}).  {\em If the increased
photoionization is the correct explanation, then we would expect
to see enhanced emission in the far infrared and submillimeter lines of
CI and [CII] compared to NGC~6822 and IC10.\/}  We will discuss
what constraints upon the physical conditions in the GMCs are imposed
by the observed line ratios in the next section.

\subsection{The Relationship Between the Super Star Clusters and the GMCs}

For comparison of the CO emission to the optical component of NGC1569,
we obtained an HST WFPC2 image from the data archive of the Space
Telescope European Coordinating Facility.
Figure~8 shows the CO 1$\rightarrow$0 contours superposed on this
image.  The F555W filter used 
corresponds approximately to the V band.  Two circles show the positions
of the SSCs, and crosses show the positions of HII regions identified
by Waller (\cite{W91}).  SSC A is near the galaxy center, and about
115 pc east of the nearest GMC, number 3.  SSC B lies
about 53 pc southeast of SSC A.  No molecular gas is directly associated
with these clusters.  This is to be expected, as these clusters represent
very strong star formation episodes approximately 15 Myr ago.  Energy
input into the ISM from the by-products of star formation (stellar winds and
supernovae) may have disrupted the natal clouds responsible
for the formation of the SSCs.  Indeed, the HI hole, the extended H$\alpha$
emission, and the hot X-ray gas all attest to the influence the SSCs have
had upon the ISM.
 
An HII region does fall partially within the contours of GMC 3, although 
it is
impossible to tell from our data if the two are physically associated.
In more massive disk galaxies, the scale heights of the cold ISM and
the young stellar populations are small enough that a spatial overlap
such as is seen here would be sufficient to assume an association
between the molecular gas and the HII region.  However, due to their 
shallower gravitational potentials, dwarf galaxies often have thicker
disks than do spirals (e.g. Holmberg II, Puche et al. \cite{PWBR}).  
Still, even if NGC~1569 has an HI scale height of $\sim$ 600 pc like Ho II,
the height of the molecular gas must necessarily be smaller.  Unless
they are in a non-equilibrium state kinematically, the ensemble of GMCs 
must lie in the plane of the galaxy.  The velocity dispersion between
the clouds (along our line of sight, of course) is $\sim$ 7 \kms, 
compared to a global $v_{fwhm}$ of 72 \kms\/ for the HI (Reakes 
\cite{R80}).  The scale height will be approximately proportional to 
the velocity dispersion (Kellman \cite{K72}), which implies a scale 
height of $\sim 60$ pc for the molecular material.  With such a low 
scale height, it is likely that GMC 3 and the HII region are physically 
close to each other.  The spatial relationship between this HII 
region and the molecular clouds is similar to what is seen in the
molecular gas south of 30 Doradus in the LMC (Johansson et al. \cite{Je98}),
where an HII region lies along the edge of the molecular gas, partially
overlapping.

Prada et al. \cite{PGM} have identified a star cluster in this region
of H$\alpha$ emission and suggested that it might be a SSC in the
process of forming.  Based upon our observations, we now consider
this to be unlikely.  The average gas density in the region near 
this cluster is $\sim$ 200 cm$^{-3}$, assuming a spherical 
geometry.  Sternberg (\cite{S98}) has argued that the density of 
molecular gas in the cloud from which the SSCs formed was $\sim 10^5$ 
cm$^{-3}$, three orders of magnitude higher than what we infer.
To recover such a high density through clumping of the molecular ISM
would require an unreasonably large volume filling factor of $\sim$ 0.001.
We note that 200 cm$^{-3}$ is only a factor of 30 larger than our
lower limit for the density of the diffuse molecular gas.

\begin{figure}
\resizebox{\hsize}{!}{\includegraphics{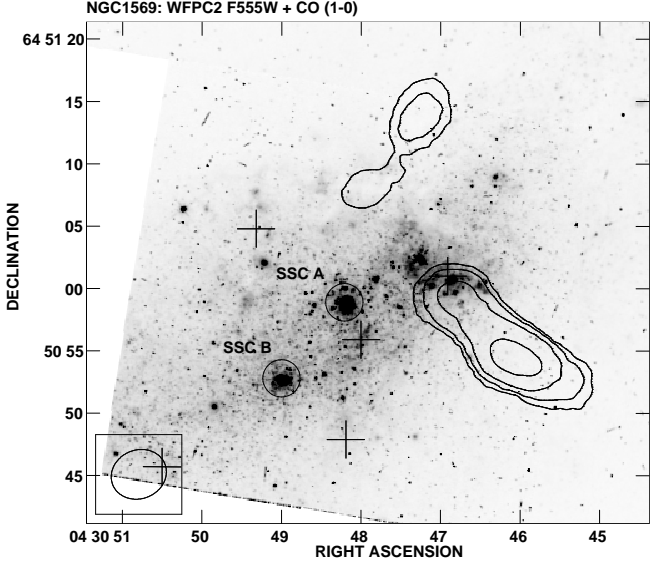}}
\caption{The contours of $^{12}$CO 1$\rightarrow$0 from Figure~3 superposed
over an HST WFPC2 image of NGC~1569 taken through the F555W filter.  The
two circles show the positions of SSCs A and B, and the crosses indicate
the positions of HII regions from Waller (\cite{W91}).
}
\label{fig8}
\end{figure}

\subsection{The Hot and Cold Phases of the ISM}

\begin{figure}
\resizebox{\hsize}{!}{\includegraphics{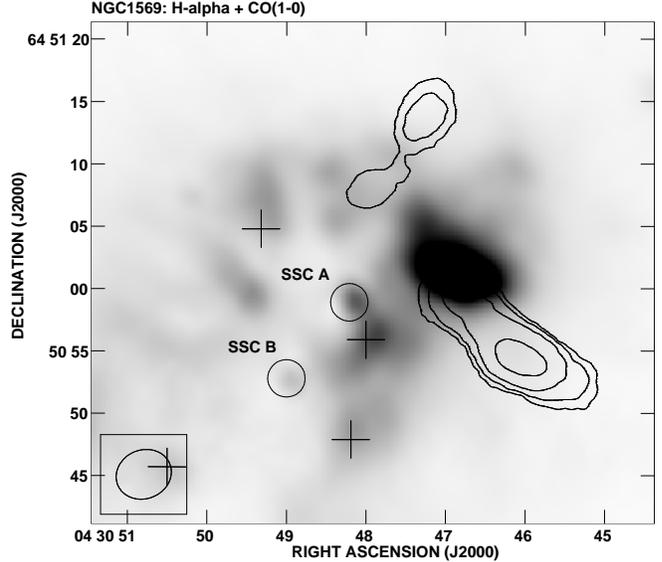}}
\caption{The contours of $^{12}$CO 1$\rightarrow$0 from Figure~3 superposed
over an H$\alpha$ image from Devost et al. 1997}
\label{fig9}
\end{figure}

NGC~1569 is often cited as a case of a dwarf galaxy experiencing a 
blowout of the ISM due to the effects of a star formation burst.
H$\alpha$ emission has been found to form a halo of emission around the
galaxy, with shell structures discernable (Hunter et al. \cite{HHG}; Devost
et al. \cite{DRD}).  The radial velocities of as much as $\pm$ 200 \kms\/
relative to the systemic velocity suggest expanding superbubbles, and X-ray 
data find hot (10$^7$ K) gas in the interior of these bubbles (Heckman et al. 
\cite{HDL}, Della Ceca et al. \cite{DCG}).  The mass of the hot X-ray gas is
1.2~$\times$~10$^6$ f$^{1\over 2}$ M$_{\sun}$, where f is the filling 
factor of the gas.  It is reasonable to expect that the violent process
of heating the gas and driving it in an outflow would leave some kind
of observable signature upon the remaining cold ISM.

The HI hole centered on SSC A discovered by Israel \& van Driel (\cite{IvD})
could be an example of this.  They argue that the data are consistent
with a picture in which the formation of the hole began about 10$^7$ yr 
ago, driven by
the expansion of supernova remnants.  The angular size they measure
for the hole is $\sim$ 10.$^{\arcsec}$, with the result that the GMCs
we have imaged lie just outside the hole.  The expanding H$\alpha$ 
bubbles have a dynamical age of $\sim$ 10$^7$ yr (Heckman et al. \cite{HDL}), 
similar to the age of the HI hole.  This provides a limit to the 
duration of the starburst that created SSC A.  Based upon the shape 
of the non-thermal radio continuum spectrum, Israel \& de Bruyn (\cite{IdB}) 
have argued that the star formation burst in NGC~1569 ended about
5~$\times$~10$^6$ yr ago, which would then suggest a burst duration of
about 5~$\times$~10$^6$.
There are three possible
scenarios regarding a connection between the hot and cold phases 
of the ISM.

\begin{enumerate}

\item The GMCs are a direct result of the expansion of the hole in
the HI driven by the hot gas.  A shell of accumulated material along
the edge of the hole might form molecular clouds.  Indeed, an expanding
shell of CO emission has been detected around the giant HII region 30
Doradus by Cohen et al. (\cite{Ce88}).  But in the case of NGC~1569, 
the GMCs do not form a shell-like distribution around the HI hole, 
nor do the kinematics of the CO emission indicate an expanding shell,
so this scenario is not very likely.  

\item Instead of
originating in a swept up shell, the GMCs may have collapsed from 
pre-existing high density material due to the shock of the outflow.
The HI maps of Israel \& van Driel do show that the peak of the HI
column density is west of the SSCs.  The lack of molecular gas elsewhere
in the vicinity of the SSCs (Greve et al. \cite{GBJM}) would be explained
if the HI elsewhere did not have a high enough density for GMCs to
collapse even with the catalyst of a passing shock.  In this case, the
GMC formation and subsequent star formation would be a prime example of
star formation being triggered by an earlier, nearby star formation
event.

\item There is no relationship.  The GMCs could predate the star formation
burst of about 15 Myr ago that created the SSCs.  Molecular clouds will
tend to form where the gas density is high.  A number of dwarf irregular
galaxies without large star formation bursts are known to have irregular,
clumpy HI distributions (e.g. Sag DIG, Young \& Lo \cite{YL}; GR8, 
Carignan et al \cite{CBF}), and near such clumps is the natural place 
to expect molecular gas.

\end{enumerate}

Option 1 is unlikely, for the reasons stated above.  Option 2 is more 
likely, given the relative geometry of the SSCs and the GMCs.  An interesting
morphological note is the H$\alpha$ arm identified by Waller (\cite{W91}).
This feature extends about 640 pc from the main part of NGC~1569 and it 
connects to the galaxy very close to the position of the GMCs and cluster
C.  A similarly shaped feature occurs in HI, sitting just to the exterior
of the H$\alpha$ feature.  Waller interprets these as the interface between
outflowing hot gas and the cool neutral material.  The GMCs would then
be positioned at the part of this interface region with the highest 
gas density.  This coincidence makes us favor option 2, although the
evidence is certainly not conclusive.
Further CO observations
of otherwise similar dwarfs which lack SSCs will help us to understand this
issue.  We discuss high resolution CO observations of other BCDs in 
section 4.4.

If we assume a star formation efficiency (SFE), defined as the fraction of 
gas mass converted into stars, we can estimate the amount of gas required 
in the burst that created the SSCs.  Sage et al. (\cite{Se92}) have
calculated {\it global} SFEs for a number of BCDs using H$\alpha$,
CO and HI observations.  These SFEs are calculated as the star formation
rate (from H$\alpha$ observations) per gas mass -- i.e. the inverse of the 
gas consumption timescale.  To
arrive at our definition of SFE, we must multiply by a burst duration, which
we will take to be 5~$\times$~10$^6$ yr.  Note that implicit in this is 
the further assumption that the current star formation rate is the same
as the average star formation rate during the burst.  We will take an average
over several BCDs, with the assumption that since they will be at different
stages in the development of their star formation episodes, the result
will be approximately an average SFE for the duration of a typical burst.
Averaging SFEs for the galaxies which most resemble NGC~1569 (which we 
define as having an HI mass within a factor of 2 of NGC~1569), we obtain 
1.9\%.  If the burst duration is longer than we have assumed, then the 
SFE will increase, because a larger fraction of the gas mass will have
been converted to stars using the average star formation rate.  For
a duration of $\sim$ 10$^8$ yr, the SFE will approach 100\%, much higher
than is expected.  However, burst durations much longer than 
5~$\times$~10$^6$ yr conflict with the dynamical age of the HI hole
(Heckman et al. \cite{HDL}) and the time for the end of the star formation
burst (Israel \& de Bruyn \cite{IdB}).

The SFE derived above compares well with the values of 2.4\% and 1.9\% 
determined
by counting individual stars in the giant HII regions NGC~595 and NGC~604 
in M33 by Wilson \& Matthews (\cite{WM}).  Of course these two are individual 
giant HII regions, and thus their SFEs are not global values, as come from 
Sage et al.  But the similarity in the SFEs between the two different
environment obtained using two different methods suggests that the values
are reasonable.

De Marchi et al. (\cite{DM}) have estimated the mass of SSC A at 
2.8~$\times$~10$^5$ M$_{\sun}$, so a SFE of 1.9\% yields an original
gas mass (H$_2$ + HI) of 1.5~$\times$~10$^7$ M$_{\sun}$.  Wilson \& 
Matthews (\cite{WM}) find the ratio of molecular to atomic hydrogen 
in NGC~595 and NGC~604 to be approximately 1:1, so this would imply
an original M$_{\rm H_2}$ of 7.5~$\times$~10$^6$ M$_{\sun}$ for the
gas that formed SSC A.  This is larger than the 1.7~$\times$~10$^6$ 
M$_{\sun}$ found in the current GMCs.  Of course there is significant
diffuse emission that was not detected in our interferometer 
observations.  In the center pointing of Greve et al. (\cite{GBJM})
this amounts to nearly a factor of 4.5, so we can estimate a lower limit
on the total H$_2$ mass in our field to be 7.7~$\times$~10$^6$.  This
is similar to the estimate for the gas that created SSC A. Thus there
is sufficient gas present now to explain the SSCs, but distributed over 
an area $\sim$ 200 pc in diameter.  This extended distribution of the 
gas may explain why we see current star formation as typical HII regions, 
but not as newly born SSCs.

%\begin{figure}
%\resizebox{\hsize}{!}{\includegraphics{XRAY.ps}}
%\caption{The contours of $^{12}$CO 1$\rightarrow$0 from Figure~3 superposed
%over an x-ray image from the HRI on ROSAT.
%}
%\label{fig9}
%\end{figure}

\subsection{Comparision with Other Galaxies}

\subsubsection{dIrrs}

Several of the irregular galaxies in the Local Group have been observed
with high spatial resolution in CO, including the LMC and SMC (each
observed with SEST), and IC10 and NGC~6822 (both observed at OVRO).  
When individual molecular clouds are
resolved, they follow a size-linewidth relationship very similar to
that of Milky Way GMCs.  The clouds seen in these more nearby galaxies 
tend to be somewhat smaller that those in NGC~1569, e.g. $\sim$ 30 pc 
diameter in the LMC and SMC (Johansson et al. \cite{Je98}; Rubio et al. 
\cite{RLB}).  It is likely that higher resolution observations would
separate the NGC~1569 clouds into smaller units, as it is known that
molecular clouds are clumpy and have a low volume filling factor.
The number of clumps contained in any arbitrary structure is not as 
important as whether or not that structure is gravitationally bound,
and in virial equilibrium.

\subsubsection{BCDs}

Except for a few cases, the history of observing CO in BCDs is largely one 
of non-detections (e.g. Young et al. \cite{Ye86}, Israel \& Burton 
\cite{IB}, Tacconi \& Young \cite{TY}, Arnault et al. \cite{ACCK}, 
Sage et al. \cite{Se92}, Israel et al. \cite{ITB}, Taylor et al. \cite{TKS} 
and Gondhalekar et al. \cite{Ge98}).  Because BCDs have comparatively high 
star formation rates, the lack of detections is not likely caused by a 
lack of molecular gas, which has been a frequent, but false, conclusion
in the past.  Instead, it is more likely to be attributed to the generally 
low metallicities of most BCDs (Searle \& Sargent \cite{SS72}) leading 
to a high CO-H$_2$ conversion factor.

Even fewer BCDs have been mapped in transitions of molecular gas.  These 
include 
NGC4214 (Becker et al. \cite{BHBW}), NGC5253 (Turner et al. \cite{TBH}), 
Henize 2-10 (Kobulnicky et al. \cite{Ke95}, Baas et al. \cite{BIK}), 
Mrk190 (Li et al. \cite{Le94}) and  III~Zw~102 (Li et al. \cite{Le93b}).
We shall discuss these in the remainder of this section to put our results
on NGC\,1569 into a larger perspective.

{\it NGC4214:}
Becker et al. (\cite{BHBW}) have mapped NGC4214 simultaneously in the 
1$\rightarrow$0 and 2$\rightarrow$0 lines of $^{12}$CO using the 
30-m telescope of IRAM.  They detected a large region of emission,
about 1000 pc $\times$ 700 pc in size, near the center of 
the galaxy.  This emission shows structure on scales of $\sim$ 500 pc,
which was approximately the resolution limit of those observations.
One feature which is well resolved has a virial mass of $\sim 10^7$
M$_{\sun}$, larger than what we have seen in NGC1569.  Clouds of
this size are rare in surveys of Galactic GMCs (e.g. Sodroski 
\cite{S91}; Sanders et al. \cite{SSS}), so perhaps these features
in NGC4214 are simply collections of unresolved smaller clouds.
When observed with a large enough beam, the GMCs we see in NGC1569
do appear as a single large cloud of diameter $\sim$ 150 - 200 pc
(Greve et al. (\cite{GBJM}).

Becker et al. find a 2$\rightarrow$0/1$\rightarrow$0 line ratio of
0.4 $\pm$ 0.1 for NGC4214.  Their data have relatively low spatial
resolution (13\arcsec\ in $^{12}$CO 2$\rightarrow$1) and NGC4214 is
more than twice as far away as NGC1569.  Therefore their line ratio
represents not a value for an individual cloud, but an average over
multiple clouds belonging to a molecular cloud complex.  In addition,
they used a single-dish telescope, so they do not have the problem
of missing flux due to a lack of short spacings.  It would be interesting
to obtain interferometric CO observations of NGC4214 in order to derive
line ratios on smaller physical scales than was possible for Becker 
et al. and see if any dense clouds with high line ratios are present,
such as we find in the case of GMC 1+2 in NGC1569.

{\it NGC5253:}
Turner et al. (\cite{TBH}) have mapped NGC5253 with OVRO at resolution of
190  $\times$ 90 pc (for the distance of 4.1 Mpc). Individual GMCs
thus are not resolved in this galaxy.  The CO distribution is only
marginally resolved at best, and is weakly detected.  Turner et al.
recover approximately one half the flux detected by single-dish
observations (e.g. Taylor et al. \cite{TKS}).  The CO emission is found 
near the optical center of the galaxy, but directly above it.  It
also lies perpendicular to the optical major axis, and along a 
dust lane.  Turner et al. suggest that the CO may have been ccreted onto
NGC5253 from another system.

{\it Henize 2-10:}
At a distance of approximately 9 Mpc, He 2-10 is too distant to resolve
even molecular complexes, as was done for NGC4214.  However, global line 
ratios have been obtained by Baas et al. (\cite{BIK}), who find a 
2$\rightarrow$1/1$\rightarrow$0 line ratio of 0.97 $\pm$ 0.16.  Because
of the low spatial resolution of their observations and the large distance
of the galaxy, this likely represents an average over different regions 
of the galaxy, with gas in different physical conditions.  Indeed, 
Baas et al. explain this line ratio, as well as the 
3$\rightarrow$2/2$\rightarrow$1 line ratio of 1.34 $\pm$ 0.17, as 
resulting from a two temperature model with a component of the CO 
emitting gas at a temperature of $<$ 10 K, and another at $>$ 75 K.  
Because we only have data in two transitions, we cannot constrain 
sophisticated models, but we certainly cannot exclude such a two 
temperature model for the CO emission in NGC1569.

The spatial distribution of $^{12}$CO 1$\rightarrow$0 emission in He 2-10 is 
described by Kobulnicky et al. (\cite{Ke95}), who obtained interferometer
observations with OVRO.  The peak of the CO emission is located 
a few arcseconds from the regions of current star formation, which
is consistent with what we find in NGC1569.  However, He 2-10 has 
an unusually extended CO distribution, with a spur of CO emitting
gas extending southeast from the star forming center of the galaxy.
This feature is also reproduced in the HI data.  Kobulnicky et al.
suggest that He 2-10 is a moderately advanced merger between two 
dwarf galaxies.  He 2-10 has a hole in the HI distribution near
the current star-forming regions, but 
unlike NGC1569, this hole is filled by the bulk of the observed 
molecular gas, and there is no evidence for an outflow of hot gas
from the starburst region. 

{\it Mrk190:} This galaxy was observed with the OVRO interferometer
in the $^{12}$CO 1$\rightarrow$0 line by Li et al. (\cite{Le94}).
Although the CO emission is only marginally resolved (distance =
17.0 Mpc), Li et al. find
evidence that the molecular gas is distributed in a ring centered on 
the galaxy center.  They suggest that starbursts in the central 
region of the galaxy have acted to clear gas out from this area,
in much the same way as is seen in NGC1569.  There is, however, 
no {\it direct}
evidence for this process in Mrk190, unlike NGC1569.  Li et al (1993a)
report single-dish observations in multiple transitions of CO for this
galaxy.  They obtain a 2$\rightarrow$1/1$\rightarrow$0 line ratio of
0.93 $\pm$ 0.25, which is roughly consistent with the line ratios for
NGC1569.  

{\it III~Zw~102:}  Despite its relatively high optical luminosity
(M$_B$ = -19.2), this galaxy is often included in studies of dwarf
galaxies (e.g. Thuan \& Martin \cite{TM81}).  Li et al. 
\cite{Le93b} obtained $^{12}$CO 1$\rightarrow$0 maps with OVRO, and
also single dish spectra in the 2$\rightarrow$1 and 1$\rightarrow$0
transitions with the IRAM 30-m telescope.  The distance to III~Zw~102
is large (23.5 Mpc), so structures like GMCs could not be resolved
in the observations.  They found a 2$\rightarrow$1/1$\rightarrow$0
line ratio of 0.66 $\pm$ 0.12, quite similar to what we found for 
GMCs 3 and 4 although, given the large errors on our line ratios,
the value for GMC 1+2 is also consistent.  This line ratio for 
III~Zw~102 is measured over the 23\arcsec\ beam of the 30-m
telescope at 115 GHz.  This corresponds to $\sim$ 2.6 kpc within
the galaxy, so the line ratio is clearly an average over a large 
number of GMCs.  The distribution of the CO emitting gas is roughly
correlated with the star formation as traced by H$\alpha$ and radio 
continuum images, but the peak of the CO emission is offset relative
to the peak of the optical emission.  This is quite similar to 
NGC1569, in which we find molecular material associated with HII regions,
but not with the SSCs.

\section{Summary and Conclusions}

We have presented CO 1$\rightarrow$0 and 2$\rightarrow$1 observations 
of the dwarf starburst galaxy NGC1569 obtained with the IRAM millimeter 
interferometer.  We confirm the result of Greve et al. (\cite{GBJM})
that the molecular gas is not associated with the super star clusters
in NGC1569, but instead with nearby HII regions.  The major results of
our study are:

\begin{enumerate}

\item The CO emission is resolved into a number of individual giant 
molecular clouds.  These GMCs have sizes and linewidths similar to 
the more massive GMCs in the Milky Way Galaxy, and to those in other
Local Group galaxies.  The 2$\rightarrow$1/1$\rightarrow$0 line ratios
of the GMCs in NGC1569 range from 0.64 $\pm$ 0.30 to 1.31 $\pm$ 0.60.
The lower values are similar to what is typically seen in the Milky 
Way GMCs, while the higher values are similar to what is observed in
starburst galaxies.

\item  The CO-H$_2$ conversion factor is found to be 6.6 $\pm$ 1.5
times the Galactic value by applying the virial theorem to 3 of the 
GMCs detected in NGC~1569.  This is approximately
three times higher than is found for NGC6822 and IC~10, two dwarf 
irregular galaxies in the Local Group with nearly the same metallicity
as NGC1569.  This difference may due to a stronger UV radiation
field in NGC1569 compared to the dwarf irregulars. Sensitive observations
of cooling lines like CI and CII, expected to be enhanced in the
presence of strong photo-dissociation, are called for to decide 
this matter. NGC1569 has 
recently experienced a strong burst of star formation which formed
the two super star clusters, and still has a number of HII regions.

\item The GMCs are observed to be just outside the edge of the HI hole
surrounding SSC A.  This hole is thought to be a region swept clear
of cold gas by the observed outflow of hot X-ray emitting gas.  It is
possible that shocks from this process may have contributed to the
formation of the GMCs, although our data do not place any constraints
on this scenario.

\end{enumerate}

\begin{acknowledgements}
We thank H. Wiesemeier (IRAM) for his assistance with the data reduction,
and D. Devost for providing us with his H$\alpha$ image of NGC~1569.
We also thank D. Bomans for interesting conversations about NGC~1569,
and L. Greggio for a detailed discussion of its star formation history.
This work has been supported by the Deutsche Forschungsgemeinschaft 
under the framework of the Graduiertenkolleg ``The Magellanic System 
and Other Dwarf Galaxies''.
\end{acknowledgements}

\end{document}